\documentclass[prl,twocolumn,superscriptaddress,amsmath,amssymb,a4,floatfix,natbib]{revtex4}

\usepackage{graphicx}% Include figure files
\usepackage{bm}% bold math
\usepackage{color}

\newcommand {\fig}[1] {Figure~\ref{#1}}   % references Figure x
\newcommand{\oneh}{$^{1}$H}
\newcommand{\Si}{$^{29}$Si}
\newcommand{\ket}[1]{\left| #1 \right\rangle}
\newcommand{\bra}[1]{\left\langle #1 \right|}
\newcommand{\mus}{$\rm{\mu}$s}

\begin{document}

\title{Magnetic field sensors using 13-spin cat states}

\author{Stephanie Simmons}
\email{stephanie.simmons@magd.ox.ac.uk}
\affiliation{Department of Materials, Oxford University, Oxford OX1 3PH, United Kingdom}

\author{Jonathan A. Jones}
\affiliation{CAESR, Clarendon Laboratory, Oxford University, Oxford OX1 3PU, United Kingdom}

\author{Steven D. Karlen}
\affiliation{Department of Materials, Oxford University, Oxford OX1 3PH, United Kingdom}

\author{Arzhang Ardavan}
\affiliation{CAESR, Clarendon Laboratory, Oxford University, Oxford OX1 3PU, United Kingdom}

\author{John~J.~L.~Morton}
\affiliation{Department of Materials, Oxford University, Oxford OX1 3PH, United Kingdom}
\affiliation{CAESR, Clarendon Laboratory, Oxford University, Oxford OX1 3PU, United Kingdom}

%(Received \today; published July 10, 2009)

\begin{abstract}
Measurement devices could benefit from entangled correlations to yield a measurement sensitivity approaching the physical Heisenberg limit. Building upon previous magnetometric work using pseudo-entangled spin states in solution-state NMR, we present two conceptual advancements to better prepare and interpret the pseudo-entanglement resource. We apply these to a 13-spin cat state to measure the local magnetic field with a 12.2 sensitivity increase over an equivalent number of isolated spins. 
\end{abstract}
\maketitle

\section{Introduction}

%%%%% NMR as a field sensor, and how entangled states evolve more quickly approaching the Heisenberg limit.
Many technologies are looking to quantum mechanics as a way to dramatically improve upon current capabilities \cite{yurke86,giovannetti04}.
As examples, interferometry \cite{quantuminf,noonz,quantuminferometry}, metrology \cite{quantummetrology}, lithography \cite{quantumlithography}, and information processing \cite{IonTrapEntSwap,KLM,graphstateEnt} are pursuing quantum information techniques to make use of the benefits of highly correlated entangled states. Even environments which normally use the pseudopure approach such as Nuclear Magnetic Resonance (NMR) \cite{Cory1997,Gershenfeld1997,pseudopure,JonesOverview} can exploit a pseudo-entanglement resource for highly sensitive magnetic field measurements \cite{cat9}.

In a locally homogeneous magnetic field, an isolated nuclear spin will precess according to its Larmor frequency which depends only on the nuclear species and the magnetic field \cite{spindy}. Consider the (unnormalised) state $\ket{0} + \ket{1}$ in a rotating frame, where the states $\ket{0}$ and $\ket{1}$ refer to the parallel and anti-parallel spin eigenstates. After a time $t$ such a state would evolve into the state $\ket{0} + \exp(i \gamma B_0 t)\ket{1}$, where the gyromagnetic ratio $\gamma$ is known, so the acquired phase can be used to deduce the local magnetic field $B_0$. A set of $N$ isolated spins can therefore serve as a microscopic magnetic field sensor \cite{spindy}, with a measurement sensitivity proportional to $\sqrt{N}$. This degree of precision is known as the `standard quantum limit' \cite{sql}.

It is possible to exceed this limit by making use of quantum entanglement.  If we assemble the $N$ spins in the state $\ket{\textbf{0}}+\ket{\textbf{1}}=\ket{00...0} + \ket{11...1}$, this will evolve, after a time $t$, into the state $\ket{\textbf{0}} + \exp(i N \gamma B_0 t)\ket{\textbf{1}}$. This evolution allows us to determine $B_0$ with an increased sensitivity compared to measuring each spin's evolution independently. The degree of sensitivity approaches the fundamental Heisenberg uncertainty relation \cite{boll96}, in that we are theoretically able to have a sensitivity proportional to $N$ using $N$ particles. In practice, the faster decoherence rates of these coherent states can reduce their sensitivity, but there will still be a net improvement in sensitivity so long as the decoherence rates scale sub-linearly with $N$ \cite{cat9} which is common in NMR \cite{Kro2004}.

We recently reported proof of principle experiments exploiting pseudo-entanglement in nuclear spin ensembles \cite{cat9}. Here we grow the size of the cat state from 10 to 13 spins and address some limitations of the previous approach. We incorporate a polarisation-priming sequence that more intelligently exploits the pseudo-entanglement resource and simplify the field estimation by disconnecting the centre spin during measurement.

\section{Sensor Selection}

%%%%% why we picked this molecule as a sensor molecule - symmetry, tumbling
To quickly generate large states such as $\ket{\textbf{0}}+\ket{\textbf{1}}$, referred to as `NOON' \cite{noonz,afek2010} or `cat' \cite{khitrin12cat,emersonentanglement,nmrcat,benchmark12,atomcat} states, we chose natural abundance solution-state tetramethylsilane (TMS) as our sensor compound. We sought a central spin-active nucleus distinct from, and surrounded by, many chemically equivalent outer spin-active nuclei, allowing us to address all peripheral spins globally; this highly symmetric configuration means that the pulse sequence complexity is independent of the number of spins, in contrast with previous work \cite{khitrin12cat,emersonentanglement,nmrcat,benchmark12}. Roughly 4.7\% of the molecules consist of one \Si\ spin surrounded by twelve \oneh\ spins (isotopic labeling could in principle be used to increase this proportion, but was not used here). Such a molecule is capable of hosting a 13-spin `cat' state.

%%%% MSSM and how fast they each should evolve, post-selecting, etc.
Reading out the thermal state on the centre \Si\ spin produces thirteen peaks corresponding to the distribution of up and down spins in the nearby hydrogen nuclei. It is convenient to assign a number, $\ell$, to each of these peaks corresponding to their `lopsidedness', that is $\ell=U-D$ where $U$ and $D$ are the number of intramolecular up and down proton spins, respectively. 

The basic sequence consists of a Hadamard gate on the centre spin to generate the state $\ket{0} + \ket{1}$, a controlled-not (CNOT) gate conditional upon the state of the centre spin, followed by an evolution delay $t$ before reversing the previous CNOT so that the final phase is mapped onto the centre spin for readout. This sequence can either be applied to a pseudopure state (corresponding to a single peak) or the entire thermal state (all thirteen peaks with a Boltzmann distribution of intensities). The two methods are computationally equivalent, but starting from the thermal state gives a stronger total readout signal. 

The internal peaks will usefully pick up phase proportional to their lopsidedness \cite{cat9}. 
The application of the Hadamard and CNOT to a given line of lopsidedness $\ell$ generates the state 
\begin{eqnarray}
\rho_{MSSM} = \sum_{i} & \left(\ket{0} \ket{M,S}_i+\ket{1} \ket{S,M}_i \right) \otimes \nonumber \\
	& \left(\bra{0} \bra{M,S}_i+\bra{1} \bra{S,M}_i \right)
\end{eqnarray}
where $\ket{M,S}$ is a number state whose indices indicate the number of proton up and down spins, respectively, and the index $i$ runs over the indistinguishable permutations of spins with a lopsidedness $\ell = M-S$. The term `Many-Some, Some-Many' (or `MSSM') was introduced \cite{cat9} to describe these states.

In this scheme the phase acquired by each `MSSM' state is given by
\begin{equation}
\phi/t =B_0\,  (\ell\gamma_{\textrm{H}} + \gamma_{\textrm{Si}})
\label{phi}
\end{equation}
where $\gamma_{\textrm{H}} = 42.577$\,MHz/T and $\gamma_{\textrm{Si}} = -8.465$\,MHz/T are the gyromagnetic ratios of \oneh\ and \Si, respectively.
As NMR experiments are most conveniently described in a rotating frame \cite{spindy} the observed phase depends not directly on the magnetic field but rather on its offset $B_0$ from some assumed nominal value. With this in mind we can calculate the phase sensitivity increase of the outermost lines of TMS to be 61.4 over an individual silicon spin and 12.2 over a single hydrogen spin. 

\begin{figure}[t] \centerline
{\includegraphics[width=3.5in]{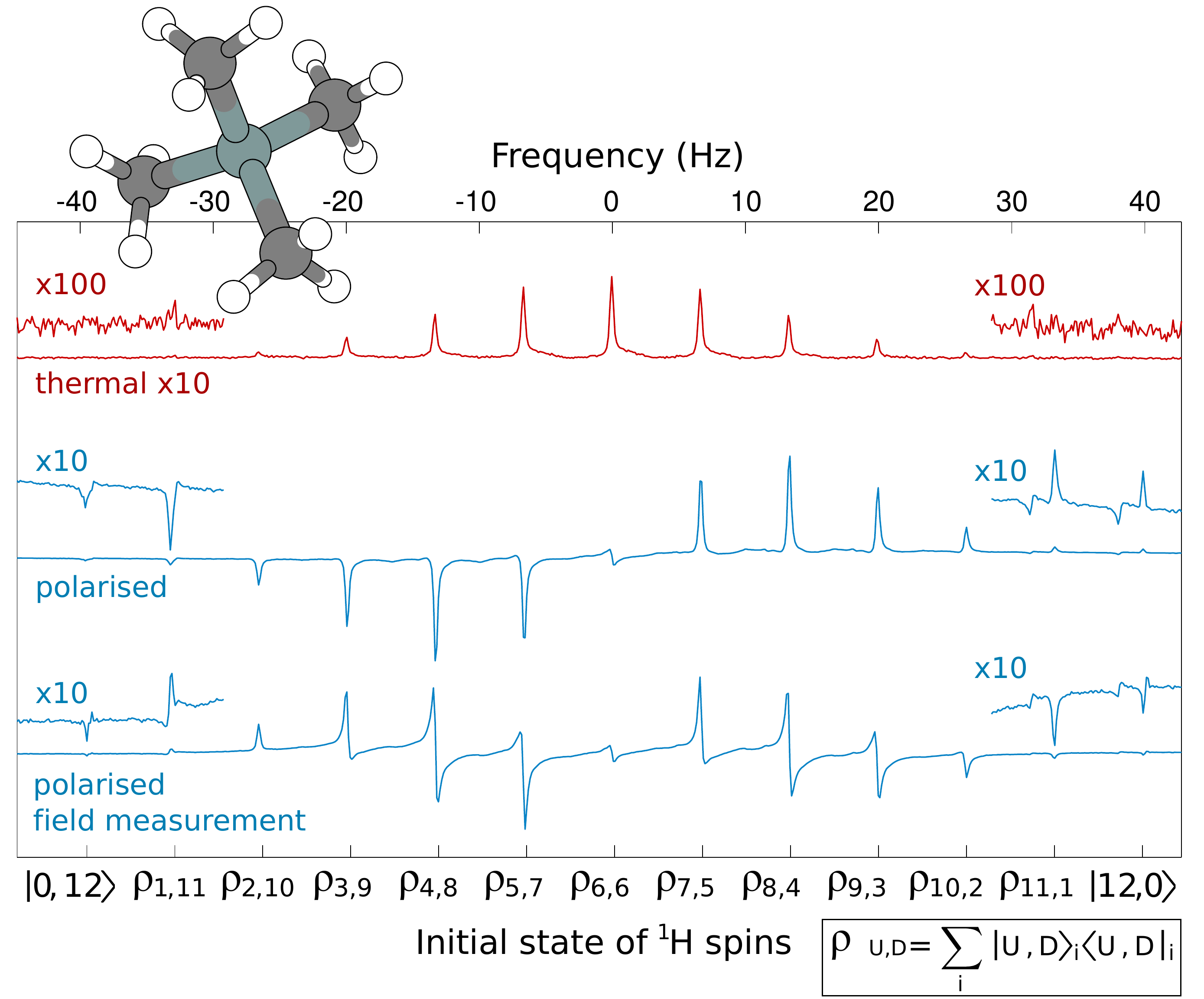}}
\caption{
(Color online) The conventional \Si\ NMR spectrum of our sensor
compoud (top spectrum, red) shows the possible transition frequencies of
the central \Si\ nucleus, which are different for the each of the 13
distinguishable states of the 12 surrounding \oneh\ nuclei,
giving rise to 13 peaks.  Frequencies are plotted as offsets from the RF
reference frequency, which is placed at the centre of the multiplet; as
in all NMR spectra the y-axis is arbitrary but spectra are plotted on a
constant scale except where expanded as indicated.  The relative
intensity of each peak depends on the Boltzmann distribution of
populations, which at high temperatures is well described by a binomial
distribution, so that the outermost peaks are too weak to be seen.  These
thermal intensities can be amplified by polarisation priming (blue
spectra), which is particularly beneficial for large sensors, allowing
the outermost peaks to be seen clearly.  Each peak, corresponding to a
particular set of \oneh\ states, can be used for field measurement (bottom
spectrum), with the outermost peaks evolving more rapidly, and hence
more sensitively, than the inner peaks.}
\label{fig:siggy} 
\end{figure}

\section{Sensor Priming}

The Boltzmann distribution of populations of the \oneh\ spins leads to weak intensities for the outermost MSSM lines which are the most sensitive to magnetic field. This could be addressed by physically \cite{Anwar2004} or computationally \cite{Baugh2005} manipulating the sensor molecule. We now describe a simple approach which uses the quantum resource more efficiently than simply averaging many measurements.

Many techniques for polarisation transfer have been developed in NMR systems \cite{Sorensen1989}; these work not by increasing the polarisation, but instead by transferring polarisation from one part of the density matrix to another where it can be more effectively used.  One simple example is a CNOT gate applied to a high-$\gamma$ nucleus controlled by a low-$\gamma$ nucleus. This transfers the population difference across a transition of the high-$\gamma$ nucleus to a low-$\gamma$ transition, effectively multiplying the polarisation of the insensitive nuclei by $\gamma_R$, the ratio of the gyromagnetic ratios of the two nuclei.  (In conventional NMR experiments this is known as Insensitive Nuclei Enhanced by Polarisation Transfer (INEPT) \cite{Sorensen1989,Morris1979}).  In our highly symmetric molecular sensor, the amplitude benefits of polarisation transfer are even greater than this ratio ($\gamma_R = -5.03$) as the CNOT has the effect of increasing the signal in a negative direction for each coupled `up' \oneh\ spin (negative because $\gamma_R$ is negative) and positive direction for each connected `down' \oneh\ spin. Explicitly, a peak of lopsidedness $\ell$ in a polarisation-primed sensor undergoes an amplitude magnification $A$ according to
\begin{equation}
A(\ell) = (1 + \gamma_R \ell) 
\label{inc}
\end{equation}
This means the outermost lines---the most sensitive components of the sensor with the poorest thermal populations---are those most amplified by the polarisation transfer. The integrated intensities of the outermost lines display an approximately 60-fold increase over the thermally-polarised measurement as expected.

\begin{figure}[t] \centerline
{\includegraphics[width=3.2in]{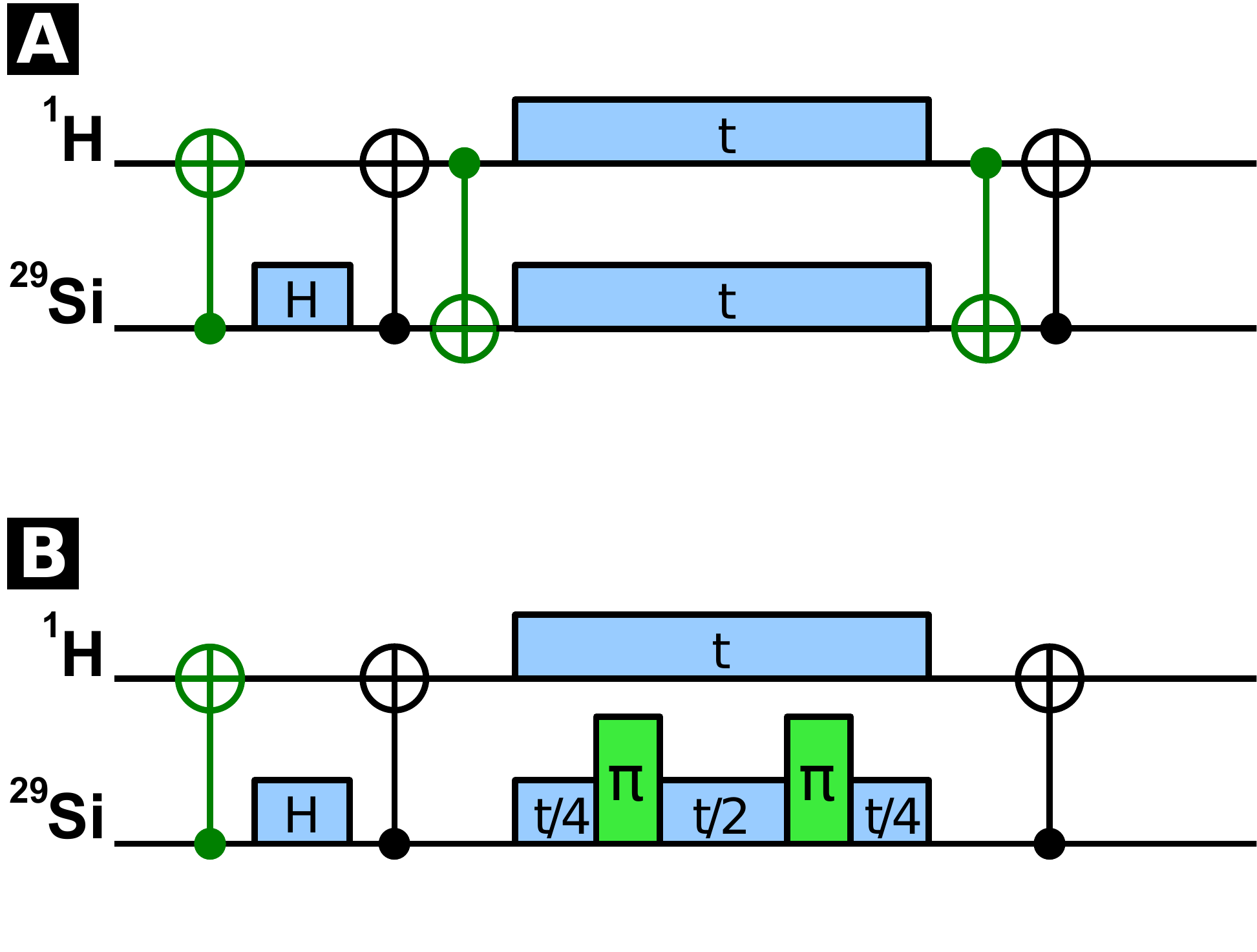}}
\caption{
(Color online) The two field sensor sequences with all improvements (coloured green) presented in this paper. Both sequences begin by polarisation-priming the state. In Sequence A, the centre two CNOT gates are `disentangling' gates designed to allow only one nuclear species (\oneh\ in TMS) to evolve during measurement. In Sequence B, two $\pi$ pulses separated by half the delay $t$ emulates a disentangling effect by reversing the phase acquired by that nucleus for half of the phase acquisition time.}
\label{fig:pulse}
\end{figure}

Polarisation-priming provides another advantage: by swapping the polarisation of sensitive and insensitive nuclei, one only needs to wait for the sensitive nuclei to rethermalise.  Although the details of relaxation processes can be complicated, in fast-tumbling spin-$1/2$ systems dominated by dipolar relaxation the $T_1$ time is inversely proportional to the square of the gyromagnetic ratio \cite{spindy}, so as the polarisation-priming sequence enhances the polarisation it also decreases the rethermalisation time. A field estimation generated from the basic sequence with polarisation priming is shown in \fig{fig:siggy}.

\section{Sensor Disentangling}

%%%DISENTANGLE
Instrumentally, there will always be some error associated with imperfect frequency detunings.  Equation~\ref{phi} assumes that the frequencies of the \oneh\ and \Si\ channels have been chosen correctly, so that they are precisely on resonance with their respective nuclei at the nominal field strength.  In general:
\begin{equation}
\phi/t =B_0\, (\ell\gamma_{\textrm{H}} + \gamma_{\textrm{Si}}) + (\ell\delta_{\textrm{H}} + \delta_{\textrm{Si}})
\label{phi2}
\end{equation}
where $\delta_{\textrm{H}}$ and $\delta_{\textrm{Si}}$ are the frequency offsets of the nuclei at the nominal field.  One can mitigate the errors generated by these terms by systematically removing them. One of these offsets (here assumed to be $\delta_{\textrm{H}}$) can be eliminated by shifting the nominal field, but it is only possible to remove both terms if the frequencies are set correctly.  An imprecisely known rotating frame offset leads to inaccurate field estimations. This requirement can be removed by `disconnecting' (disentangling)  the centre spin during the phase acquisition delay. Two methods for achieving this are introduced in \fig{fig:pulse}.

%%%%%PULSESEQUENCE

Consider how sequence A acts upon the leftmost line. In the pseudopure approximation \cite{pseudopure} the leftmost line in its thermal state is represented as $\ket{0}_{\textrm{Si}} \ket{0}_{\textrm{H}}^{\otimes 12}$, where $\otimes$ indicates a tensor product. A Hadamard gate followed by a CNOT gate conditional upon the silicon nucleus transforms the initial state into $\ket{0}_{\textrm{Si}} \ket{0}_{\textrm{H}}^{\otimes 12} + \ket{1}_{\textrm{Si}}\ket{1}_{\textrm{H}}^{\otimes 12}$. It is here that we can disentangle the central \Si\ spin from our large cat state by applying a NOT gate to \Si\ in the second term, giving
\begin{equation}
\ket{0}_{\textrm{Si}} \ket{0}_{\textrm{H}}^{\otimes 12} + \ket{0}_{\textrm{Si}}\ket{1}_{\textrm{H}}^{\otimes 12}=\ket{0}_{\textrm{Si}}\left(\ket{0}_{\textrm{H}}^{\otimes 12} + \ket{1}_{\textrm{H}}^{\otimes 12}\right)
\end{equation}
so that only the \oneh\ spins will acquire field-dependent phases.

It might seem that this approach would require a multiply-controlled NOT gate (a generalised Toffoli gate), but with a pseudopure state this is not required. It is only necessary to apply a NOT gate to the second term and \textit{not} to the first term.  This can be achieved with a modified CNOT gate \cite{JonesOverview}, with the evolution time chosen to match the separation between the outermost lines in the \Si\ multiplet rather than the conventional coupling size.

\begin{figure}[t] \centerline
{\includegraphics[width=3.5in]{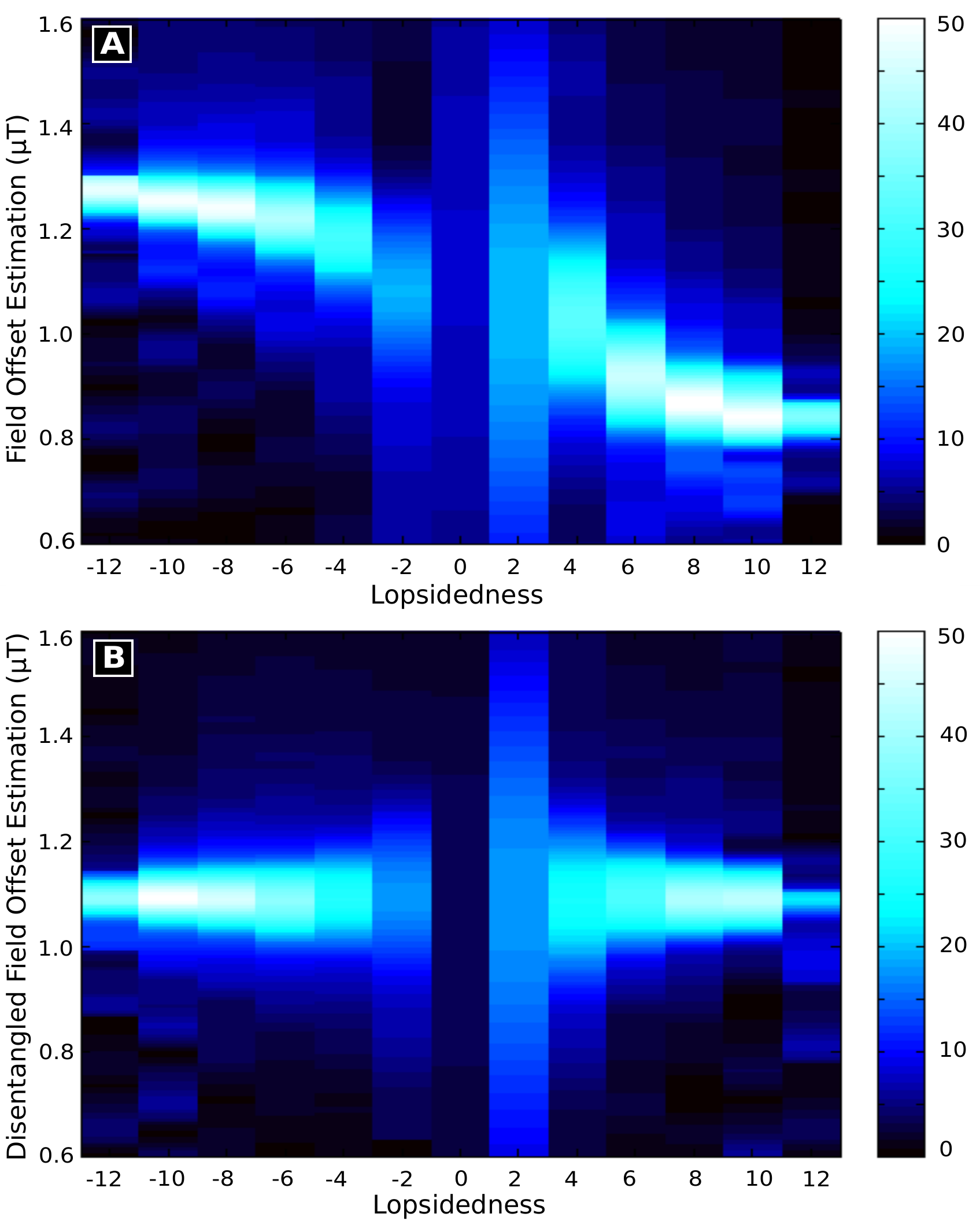}}
\caption{
(Color online) Each peak in the \Si\ NMR spectrum(labeled according to lopsidedness) can estimate the local magnetic field. The precision of each peak's estimation scales according to its absolute lopsidedness, leading to the most precise estimations at the outermost peaks. The colour amplitude chosen reflects the gain in signal from the polarisation-priming sequence; a value of 1 on each peak estimation represents its thermal binomial amplitude (See Equation \ref{inc}).
A: The field estimation is visibly sensitive to a small nonzero detuning (3.5 Hz) on the \Si\ spin if it is not disentangled during measurement.
B: The same \Si\ detuning does not distort the field estimation when applying disentangling sequence B (see \fig{fig:pulse}B) }
\label{fig:inept}
\end{figure}

This simple approach must be modified to work simultaneously with a general set of MSSM lines.  With an odd number of \oneh\ spins, this can be achieved with a conventional CNOT gate (see \fig{fig:pulse}A), which disentangles every MSSM line. This approach does not work for systems with an even number of \oneh\ spins. An alternative, simpler method (\fig{fig:pulse}B) uses echoes to refocus the inner \Si\ spin rather than disentangling it, and this can be applied to both even and odd systems. 
Under both sequences we read out the acquired phase by applying the sequence in reverse, and observing the central \Si\ spin.

\begin{figure}[t] \centerline
{\includegraphics[width=3.5in]{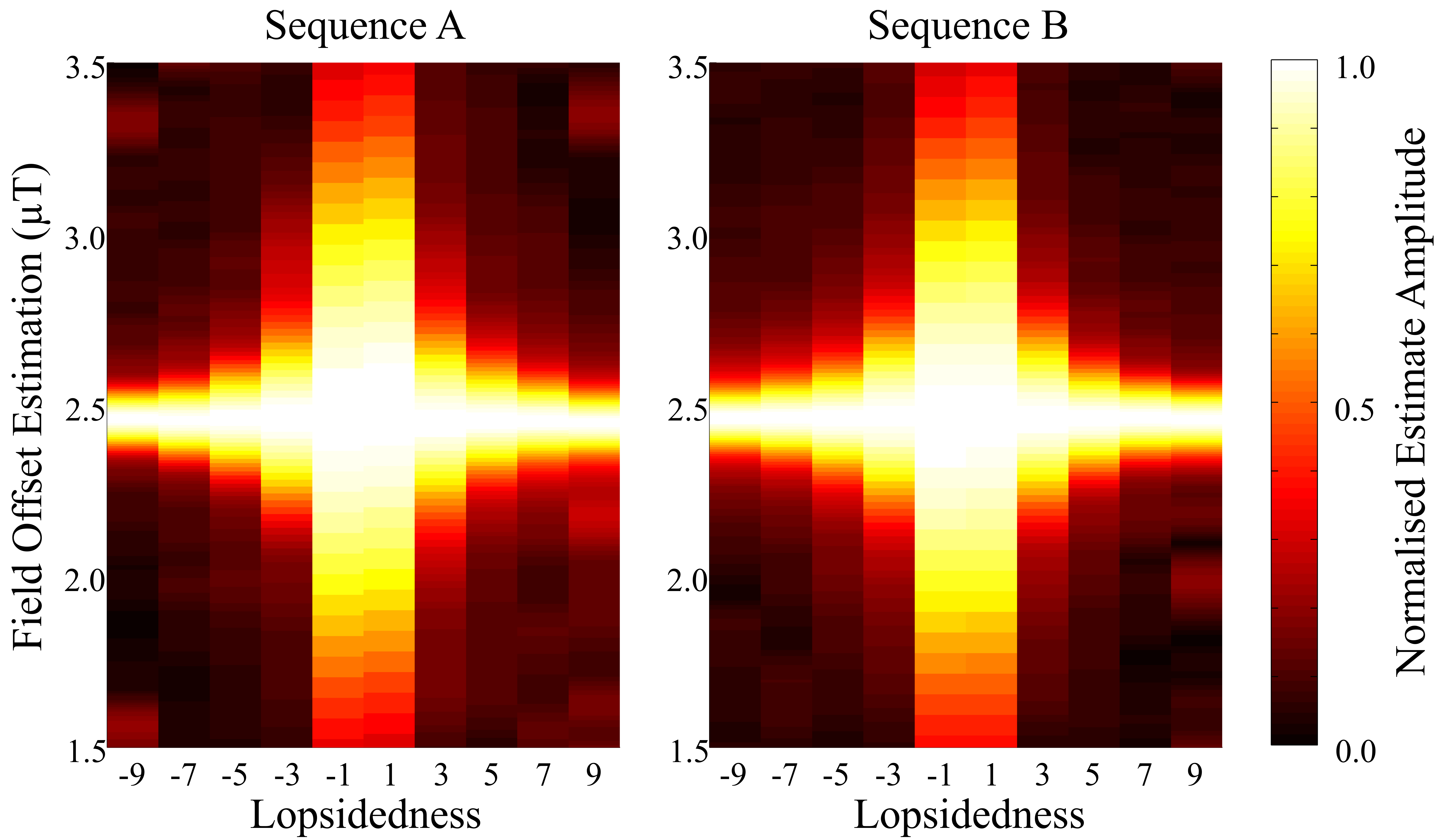}} \caption{(Color online) A comparison of the normalised field estimations from the disentangling sequences A and B as introduced in \fig{fig:pulse}. The molecular sensor used was trimethylphosphite (TMP) because of its odd number of \oneh\ nuclei.} \label{fig:compareSeqs}
\end{figure}

\begin{figure}[t] \centerline
{\includegraphics[width=3.5in]{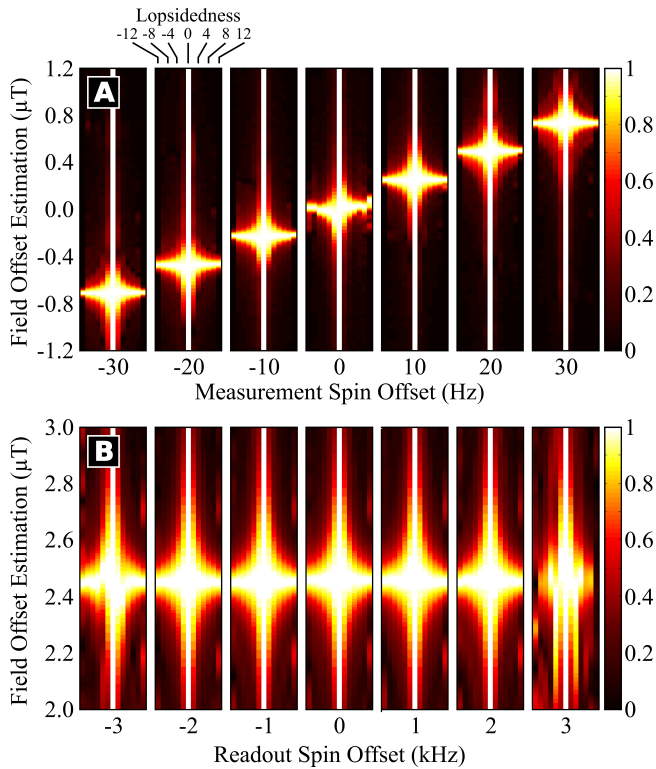}} \caption{(Color online) Field estimations using Sequence B under different conditions. A: Varying the \oneh\ measurement spin offset produces the expected field offset estimations, showing no distortions under a constant \Si\ offset of 15Hz. The data at zero offset is contaminated by low frequency artefacts arising from a number of sources. Such low frequency artefacts can sometimes be suppressed, such as the removal of axial peaks in two-dimensional NMR spectra by phase cycling \cite{wider}.  In general, however, it is preferable to measure with a slight offset so that the signal of interest is displaced from these erroneous low frequency terms.
B: A selection of disentangled field offset estimations under various \Si\ measurement spin detunings. The bandwidth of the \Si\ pulses can account for the distortions seen at offsets greater than $\pm 2$kHz. \label{fig:offsets}}
\end{figure}

To test this approach, we applied a small offset to the \Si\ channel and implemented a full field estimation with the original pulse sequence and with the modified sequence B.  As shown in Fig.~\ref{fig:inept}, the field estimation now gives different results for different lines in the multiplet if the original sequence is used, but these imperfections are removed by the modified sequence.  Sequences A and B were both successfully implemented on our original odd spin system, trimethylphosphite (See \fig{fig:compareSeqs}).  We then repeated the phase estimation with a wide range of silicon channel detunings (See \fig{fig:offsets}) and obtained indistinguishable field estimations. 

%%%%%FIELD ESTIMATION

%%DRAWBACKS, ECNOT VS RCNOT
The most obvious drawback of these new sequences is a mild sensitivity decrease from $(12\gamma_{H} + \gamma_{Si})/\gamma_{H}$ to $12$ times that of a single \oneh\ spin. Because we remove the need to accurately measure frequency offsets, however, simplicity (and potentially accuracy) is enhanced. It is important to choose the correct disentangling sequence so that all MSSM peaks are measured simultaneously, for reasons that will now be discussed.

\section{Conclusion}

%%%%ALIASING
We can extract many times more information with a single scan by considering all the peaks in the spectrum. A single NOON state sensor can easily encounter aliasing problems; in effect one already needs to know the approximate field offset to be certain of the results. A full arsenal of MSSM states provides a mechanism to avoid such problems. On an quantum sensor with N outer spins, a $\phi$ phase rotation on the outermost peak could only be aliased with a rotation of $\phi + 2 k N \pi$ for some integer k. In quantum interferometric terms \cite{durkin}, such a sensor simultaneously displays both local and global phase distinguishability. Such anti-aliasing effects are a desirable property of these highly symmetric sensor molecules. 

%%%%where are we going
For even more sensitive measurement, larger sensors can be employed, potentially with iterative (or other) geometries. To extract information from the outermost peaks of such very large sensors, polarisation amplification methods such as Dynamic Nuclear Polarisation \cite{overhauser} can be applied in addition to the methods outlined above. All sensors can benefit from both the simplified field estimation afforded by disentanglement methods, and from polarisation-priming the pseudo-entanglement resource as introduced in this manuscript.

In conclusion, we have generated 13-spin pseudo-`cat' states for entanglement-enhanced magnetometry. We have proposed and applied innovations to improve the stability and resolution of the entanglement resource. Errors arising from imperfect knowledge of system variables are removed by two different disentanglement methods, and the overall weighted enhancement afforded by polarisation-priming shows approximately a 60-fold increase in the most sensitive components of the sensor. 

This research is supported by the EPSRC through the QIP IRC www.qipirc.org (GR/S82176/01) and CAESR (EP/D048559/1). We thank Vasileia Filidou for graphics assistance. S.S. thanks Magdalen College, Oxford. A.A. and J.J.L.M. thank the Royal Society.

\section{Methods}

The sample was a 1:1 by volume solution of tetra\-methyl\-silane and acetone-$\textrm{d}_6$, degassed using freeze-pump-thaw cycles, and flame sealed in a 5~mm Wilmad LabGlass NMR tube. All NMR experiments were performed at a temperature of $20^\circ$C on a Varian INOVA 600 spectrometer using a broadband tunable X\{H\} probe with a $^2$H~lock with a 4-step phase cycle to cancel receiver errors. $\pi/2$ pulse lengths were approximately 27~\mus~on the hydrogen channel and 17~\mus~ on the silicon channel. The spin--spin coupling ${}^{3}J_{\textrm{SiH}}$ was 6.63\,Hz.  Measured ${}^{29}\textrm{Si}$ relaxation times were $\textrm{T}_2=1.2\,\textrm{s}$ and $\textrm{T}_1=25.4\,\textrm{s}$, while \oneh\ relaxation times were $\textrm{T}_2=1.6\,\textrm{s}$ and $\textrm{T}_1=8.9\,\textrm{s}$. The measured $\textrm{T}^*_2$ times for the \oneh\ spin and NOON state were $0.37\,\textrm{s}$ and $0.28\,\textrm{s}$, respectively.

Quantum logic gates were implemented using standard NMR techniques \cite{JonesOverview}. Hadamard gates were applied as $\pi/2_{-y} \pi_{z}$. C-NOT gates, equivalent to a controlled-phase gate surrounded by Hadamard gates on one channel \cite{JonesOverview}, were implemented as two \oneh\ $\pi/2$ pulses separated by a spin echo of length
$1/2J$, where $J$ is the spin--spin coupling constant. All $Z$ gates were realised as phase shifts in the pulses that followed \cite{JonesOverview}. To reduce off-resonance and RF inhomogeneity errors, spin-echoes were constructed with two simultaneous $\pi_{x}$ pulses at times $1/8J$ and $3/8J$, and all pulses were implemented as simultaneous, equal-duration BB1 composite pulses \cite{Cummins2003}. Implementing such pulses used suitable amplitude adjustments and `$0$-degree' identity gate pulses where required.

Data was apodised with a Hamming filter and Fourier transformed using matNMR \cite{matNMR} version 3.9.59. The spectra with no phase accumulation delay was phased, and that phase correction was applied to all other spectra for consistency. Spectra were then exported to Matlab for final processing. 

\bibliography{bib}

\begin{thebibliography}{34}
\expandafter\ifx\csname natexlab\endcsname\relax\def\natexlab#1{#1}\fi
\expandafter\ifx\csname bibnamefont\endcsname\relax
  \def\bibnamefont#1{#1}\fi
\expandafter\ifx\csname bibfnamefont\endcsname\relax
  \def\bibfnamefont#1{#1}\fi
\expandafter\ifx\csname citenamefont\endcsname\relax
  \def\citenamefont#1{#1}\fi
\expandafter\ifx\csname url\endcsname\relax
  \def\url#1{\texttt{#1}}\fi
\expandafter\ifx\csname urlprefix\endcsname\relax\def\urlprefix{URL }\fi
\providecommand{\bibinfo}[2]{#2}
\providecommand{\eprint}[2][]{\url{#2}}

\bibitem[{\citenamefont{{Yurke}}(1986)}]{yurke86}
\bibinfo{author}{\bibfnamefont{B.}~\bibnamefont{{Yurke}}},
  \bibinfo{journal}{Phys. Rev. Lett.} \textbf{\bibinfo{volume}{56}},
  \bibinfo{pages}{1515} (\bibinfo{year}{1986}).

\bibitem[{\citenamefont{{Giovannetti} et~al.}(2004)\citenamefont{{Giovannetti},
  {Lloyd}, and {Maccone}}}]{giovannetti04}
\bibinfo{author}{\bibfnamefont{V.}~\bibnamefont{{Giovannetti}}},
  \bibinfo{author}{\bibfnamefont{S.}~\bibnamefont{{Lloyd}}}, \bibnamefont{and}
  \bibinfo{author}{\bibfnamefont{L.}~\bibnamefont{{Maccone}}},
  \bibinfo{journal}{Science} \textbf{\bibinfo{volume}{306}},
  \bibinfo{pages}{1330} (\bibinfo{year}{2004}).

\bibitem[{\citenamefont{Holland and Burnett}(1993)}]{quantuminf}
\bibinfo{author}{\bibfnamefont{M.~J.} \bibnamefont{Holland}} \bibnamefont{and}
  \bibinfo{author}{\bibfnamefont{K.}~\bibnamefont{Burnett}},
  \bibinfo{journal}{Phys. Rev. Lett.} \textbf{\bibinfo{volume}{71}},
  \bibinfo{pages}{1355} (\bibinfo{year}{1993}).

\bibitem[{\citenamefont{{Walther} et~al.}(2004)\citenamefont{{Walther}, {Pan},
  {Aspelmeyer}, {Ursin}, {Gasparoni}, and {Zeilinger}}}]{noonz}
\bibinfo{author}{\bibfnamefont{P.}~\bibnamefont{{Walther}}},
  \bibinfo{author}{\bibfnamefont{J.-W.} \bibnamefont{{Pan}}},
  \bibinfo{author}{\bibfnamefont{M.}~\bibnamefont{{Aspelmeyer}}},
  \bibinfo{author}{\bibfnamefont{R.}~\bibnamefont{{Ursin}}},
  \bibinfo{author}{\bibfnamefont{S.}~\bibnamefont{{Gasparoni}}},
  \bibnamefont{and}
  \bibinfo{author}{\bibfnamefont{A.}~\bibnamefont{{Zeilinger}}},
  \bibinfo{journal}{Nature} \textbf{\bibinfo{volume}{429}},
  \bibinfo{pages}{158} (\bibinfo{year}{2004}).

\bibitem[{\citenamefont{Mitchell et~al.}(2004)\citenamefont{Mitchell, Lundeen,
  and Steinberg}}]{quantuminferometry}
\bibinfo{author}{\bibfnamefont{M.~W.} \bibnamefont{Mitchell}},
  \bibinfo{author}{\bibfnamefont{J.~S.} \bibnamefont{Lundeen}},
  \bibnamefont{and} \bibinfo{author}{\bibfnamefont{A.~M.}
  \bibnamefont{Steinberg}}, \bibinfo{journal}{Nature}
  \textbf{\bibinfo{volume}{429}}, \bibinfo{pages}{161} (\bibinfo{year}{2004}).

\bibitem[{\citenamefont{Kok et~al.}(2004)\citenamefont{Kok, Braunstein, and
  Dowling}}]{quantummetrology}
\bibinfo{author}{\bibfnamefont{P.}~\bibnamefont{Kok}},
  \bibinfo{author}{\bibfnamefont{S.~L.} \bibnamefont{Braunstein}},
  \bibnamefont{and} \bibinfo{author}{\bibfnamefont{J.~P.}
  \bibnamefont{Dowling}}, \bibinfo{journal}{Journal of Optics B: Quant.
  Semiclass. Opt.} \textbf{\bibinfo{volume}{6}}, \bibinfo{pages}{S811}
  (\bibinfo{year}{2004}).

\bibitem[{\citenamefont{Boto et~al.}(2000)\citenamefont{Boto, Kok, Abrams,
  Braunstein, Williams, and Dowling}}]{quantumlithography}
\bibinfo{author}{\bibfnamefont{A.~N.} \bibnamefont{Boto}},
  \bibinfo{author}{\bibfnamefont{P.}~\bibnamefont{Kok}},
  \bibinfo{author}{\bibfnamefont{D.~S.} \bibnamefont{Abrams}},
  \bibinfo{author}{\bibfnamefont{S.~L.} \bibnamefont{Braunstein}},
  \bibinfo{author}{\bibfnamefont{C.~P.} \bibnamefont{Williams}},
  \bibnamefont{and} \bibinfo{author}{\bibfnamefont{J.~P.}
  \bibnamefont{Dowling}}, \bibinfo{journal}{Phys. Rev. Lett.}
  \textbf{\bibinfo{volume}{85}}, \bibinfo{pages}{2733} (\bibinfo{year}{2000}).

\bibitem[{\citenamefont{Riebe et~al.}(2008)\citenamefont{Riebe, Monz, Kim,
  Villar, Schindler, Chwalla, Hennrich, and Blatt}}]{IonTrapEntSwap}
\bibinfo{author}{\bibfnamefont{M.}~\bibnamefont{Riebe}},
  \bibinfo{author}{\bibfnamefont{T.}~\bibnamefont{Monz}},
  \bibinfo{author}{\bibfnamefont{K.}~\bibnamefont{Kim}},
  \bibinfo{author}{\bibfnamefont{A.~S.} \bibnamefont{Villar}},
  \bibinfo{author}{\bibfnamefont{P.}~\bibnamefont{Schindler}},
  \bibinfo{author}{\bibfnamefont{M.}~\bibnamefont{Chwalla}},
  \bibinfo{author}{\bibfnamefont{M.}~\bibnamefont{Hennrich}}, \bibnamefont{and}
  \bibinfo{author}{\bibfnamefont{R.}~\bibnamefont{Blatt}},
  \bibinfo{journal}{Nature} \textbf{\bibinfo{volume}{4}}, \bibinfo{pages}{839}
  (\bibinfo{year}{2008}).

\bibitem[{\citenamefont{Knill et~al.}(2001)\citenamefont{Knill, Laflamme, and
  Milburn}}]{KLM}
\bibinfo{author}{\bibfnamefont{E.}~\bibnamefont{Knill}},
  \bibinfo{author}{\bibfnamefont{R.}~\bibnamefont{Laflamme}}, \bibnamefont{and}
  \bibinfo{author}{\bibfnamefont{G.~J.} \bibnamefont{Milburn}},
  \bibinfo{journal}{Nature} \textbf{\bibinfo{volume}{409}}, \bibinfo{pages}{46}
  (\bibinfo{year}{2001}).

\bibitem[{\citenamefont{Hein et~al.}(2004)\citenamefont{Hein, Eisert, and
  Briegel}}]{graphstateEnt}
\bibinfo{author}{\bibfnamefont{M.}~\bibnamefont{Hein}},
  \bibinfo{author}{\bibfnamefont{J.}~\bibnamefont{Eisert}}, \bibnamefont{and}
  \bibinfo{author}{\bibfnamefont{H.~J.} \bibnamefont{Briegel}},
  \bibinfo{journal}{Phys. Rev. A} \textbf{\bibinfo{volume}{69}},
  \bibinfo{pages}{062311} (\bibinfo{year}{2004}).

\bibitem[{\citenamefont{Cory et~al.}(1997)\citenamefont{Cory, Fahmy, and
  Havel}}]{Cory1997}
\bibinfo{author}{\bibfnamefont{D.~G.} \bibnamefont{Cory}},
  \bibinfo{author}{\bibfnamefont{A.~F.} \bibnamefont{Fahmy}}, \bibnamefont{and}
  \bibinfo{author}{\bibfnamefont{T.~F.} \bibnamefont{Havel}},
  \bibinfo{journal}{Proc. Natl. Acad. Sci. USA} \textbf{\bibinfo{volume}{94}},
  \bibinfo{pages}{1634} (\bibinfo{year}{1997}).

\bibitem[{\citenamefont{Gershenfeld and Chuang}(1997)}]{Gershenfeld1997}
\bibinfo{author}{\bibfnamefont{N.~A.} \bibnamefont{Gershenfeld}}
  \bibnamefont{and} \bibinfo{author}{\bibfnamefont{I.~L.}
  \bibnamefont{Chuang}}, \bibinfo{journal}{Science}
  \textbf{\bibinfo{volume}{275}}, \bibinfo{pages}{350} (\bibinfo{year}{1997}).

\bibitem[{\citenamefont{{Knill} et~al.}(1998)\citenamefont{{Knill}, {Chuang},
  and {Laflamme}}}]{pseudopure}
\bibinfo{author}{\bibfnamefont{E.}~\bibnamefont{{Knill}}},
  \bibinfo{author}{\bibfnamefont{I.}~\bibnamefont{{Chuang}}}, \bibnamefont{and}
  \bibinfo{author}{\bibfnamefont{R.}~\bibnamefont{{Laflamme}}},
  \bibinfo{journal}{Phys. Rev. A.} \textbf{\bibinfo{volume}{57}},
  \bibinfo{pages}{3348} (\bibinfo{year}{1998}).

\bibitem[{\citenamefont{Jones}(2001)}]{JonesOverview}
\bibinfo{author}{\bibfnamefont{J.~A.} \bibnamefont{Jones}},
  \bibinfo{journal}{Prog. Nucl. Magn. Reson. Spectrosc}
  \textbf{\bibinfo{volume}{38}}, \bibinfo{pages}{325 } (\bibinfo{year}{2001}).

\bibitem[{\citenamefont{Jones et~al.}(2009)\citenamefont{Jones, Karlen,
  Fitzsimons, Ardavan, Benjamin, Briggs, and Morton}}]{cat9}
\bibinfo{author}{\bibfnamefont{J.~A.} \bibnamefont{Jones}},
  \bibinfo{author}{\bibfnamefont{S.~D.} \bibnamefont{Karlen}},
  \bibinfo{author}{\bibfnamefont{J.}~\bibnamefont{Fitzsimons}},
  \bibinfo{author}{\bibfnamefont{A.}~\bibnamefont{Ardavan}},
  \bibinfo{author}{\bibfnamefont{S.~C.} \bibnamefont{Benjamin}},
  \bibinfo{author}{\bibfnamefont{G.~A.~D.} \bibnamefont{Briggs}},
  \bibnamefont{and} \bibinfo{author}{\bibfnamefont{J.~J.~L.}
  \bibnamefont{Morton}}, \bibinfo{journal}{Science}
  \textbf{\bibinfo{volume}{324}}, \bibinfo{pages}{1166} (\bibinfo{year}{2009}).

\bibitem[{\citenamefont{Levitt}(2001)}]{spindy}
\bibinfo{author}{\bibfnamefont{M.~H.} \bibnamefont{Levitt}},
  \emph{\bibinfo{title}{Spin Dynamics: Basics of Nuclear Magnetic Resonance;
  2nd ed.}} (\bibinfo{publisher}{Wiley}, \bibinfo{address}{Chichester},
  \bibinfo{year}{2001}).

\bibitem[{\citenamefont{{Luis} and {S{\'a}nchez-Soto}}(1992)}]{sql}
\bibinfo{author}{\bibfnamefont{A.}~\bibnamefont{{Luis}}} \bibnamefont{and}
  \bibinfo{author}{\bibfnamefont{L.~L.} \bibnamefont{{S{\'a}nchez-Soto}}},
  \bibinfo{journal}{Opt. Comm.} \textbf{\bibinfo{volume}{89}},
  \bibinfo{pages}{140} (\bibinfo{year}{1992}).

\bibitem[{\citenamefont{Bollinger et~al.}(1996)\citenamefont{Bollinger, Itano,
  Wineland, and Heinzen}}]{boll96}
\bibinfo{author}{\bibfnamefont{J.~J.~.} \bibnamefont{Bollinger}},
  \bibinfo{author}{\bibfnamefont{W.~M.} \bibnamefont{Itano}},
  \bibinfo{author}{\bibfnamefont{D.~J.} \bibnamefont{Wineland}},
  \bibnamefont{and} \bibinfo{author}{\bibfnamefont{D.~J.}
  \bibnamefont{Heinzen}}, \bibinfo{journal}{Phys. Rev. A}
  \textbf{\bibinfo{volume}{54}}, \bibinfo{pages}{R4649} (\bibinfo{year}{1996}).

\bibitem[{\citenamefont{Krojanski and Suter}(2004)}]{Kro2004}
\bibinfo{author}{\bibfnamefont{H.~G.} \bibnamefont{Krojanski}}
  \bibnamefont{and} \bibinfo{author}{\bibfnamefont{D.}~\bibnamefont{Suter}},
  \bibinfo{journal}{\prl} \textbf{\bibinfo{volume}{93}},
  \bibinfo{pages}{090501} (\bibinfo{year}{2004}).

\bibitem[{\citenamefont{{Afek} et~al.}(2010)\citenamefont{{Afek}, {Ambar}, and
  {Silberberg}}}]{afek2010}
\bibinfo{author}{\bibfnamefont{I.}~\bibnamefont{{Afek}}},
  \bibinfo{author}{\bibfnamefont{O.}~\bibnamefont{{Ambar}}}, \bibnamefont{and}
  \bibinfo{author}{\bibfnamefont{Y.}~\bibnamefont{{Silberberg}}},
  \bibinfo{journal}{Science} \textbf{\bibinfo{volume}{328}},
  \bibinfo{pages}{879} (\bibinfo{year}{2010}).

\bibitem[{\citenamefont{Lee and Khitrin}(2005)}]{khitrin12cat}
\bibinfo{author}{\bibfnamefont{J.~S.} \bibnamefont{Lee}} \bibnamefont{and}
  \bibinfo{author}{\bibfnamefont{A.~K.} \bibnamefont{Khitrin}},
  \bibinfo{journal}{App. Phys. Lett.} \textbf{\bibinfo{volume}{87}},
  \bibinfo{pages}{204109} (\bibinfo{year}{2005}).

\bibitem[{\citenamefont{Cappellaro et~al.}(2005)\citenamefont{Cappellaro,
  Emerson, Boulant, Ramanathan, Lloyd, and Cory}}]{emersonentanglement}
\bibinfo{author}{\bibfnamefont{P.}~\bibnamefont{Cappellaro}},
  \bibinfo{author}{\bibfnamefont{J.}~\bibnamefont{Emerson}},
  \bibinfo{author}{\bibfnamefont{N.}~\bibnamefont{Boulant}},
  \bibinfo{author}{\bibfnamefont{C.}~\bibnamefont{Ramanathan}},
  \bibinfo{author}{\bibfnamefont{S.}~\bibnamefont{Lloyd}}, \bibnamefont{and}
  \bibinfo{author}{\bibfnamefont{D.~G.} \bibnamefont{Cory}},
  \bibinfo{journal}{Phys. Rev. Lett.} \textbf{\bibinfo{volume}{94}},
  \bibinfo{pages}{020502} (\bibinfo{year}{2005}).

\bibitem[{\citenamefont{Knill et~al.}(2000)\citenamefont{Knill, Laflamme,
  Martinez, and TSeng}}]{nmrcat}
\bibinfo{author}{\bibfnamefont{E.}~\bibnamefont{Knill}},
  \bibinfo{author}{\bibfnamefont{R.}~\bibnamefont{Laflamme}},
  \bibinfo{author}{\bibfnamefont{R.}~\bibnamefont{Martinez}}, \bibnamefont{and}
  \bibinfo{author}{\bibfnamefont{C.-H.} \bibnamefont{TSeng}},
  \bibinfo{journal}{Nature} \textbf{\bibinfo{volume}{404}},
  \bibinfo{pages}{368} (\bibinfo{year}{2000}).

\bibitem[{\citenamefont{{Negrevergne} et~al.}(2006)\citenamefont{{Negrevergne},
  {Mahesh}, {Ryan}, {Ditty}, {Cyr-Racine}, {Power}, {Boulant}, {Havel}, {Cory},
  and {Laflamme}}}]{benchmark12}
\bibinfo{author}{\bibfnamefont{C.}~\bibnamefont{{Negrevergne}}},
  \bibinfo{author}{\bibfnamefont{T.~S.} \bibnamefont{{Mahesh}}},
  \bibinfo{author}{\bibfnamefont{C.~A.} \bibnamefont{{Ryan}}},
  \bibinfo{author}{\bibfnamefont{M.}~\bibnamefont{{Ditty}}},
  \bibinfo{author}{\bibfnamefont{F.}~\bibnamefont{{Cyr-Racine}}},
  \bibinfo{author}{\bibfnamefont{W.}~\bibnamefont{{Power}}},
  \bibinfo{author}{\bibfnamefont{N.}~\bibnamefont{{Boulant}}},
  \bibinfo{author}{\bibfnamefont{T.}~\bibnamefont{{Havel}}},
  \bibinfo{author}{\bibfnamefont{D.~G.} \bibnamefont{{Cory}}},
  \bibnamefont{and}
  \bibinfo{author}{\bibfnamefont{R.}~\bibnamefont{{Laflamme}}},
  \bibinfo{journal}{Phys. Rev. Lett.} \textbf{\bibinfo{volume}{96}},
  \bibinfo{pages}{170501} (\bibinfo{year}{2006}).

\bibitem[{\citenamefont{Leibfried et~al.}(2005)\citenamefont{Leibfried, Knill,
  Seidelin, Britton, Blakestad, Chiaverini, Hume, Itano, Jost, Langer
  et~al.}}]{atomcat}
\bibinfo{author}{\bibfnamefont{D.}~\bibnamefont{Leibfried}},
  \bibinfo{author}{\bibfnamefont{E.}~\bibnamefont{Knill}},
  \bibinfo{author}{\bibfnamefont{S.}~\bibnamefont{Seidelin}},
  \bibinfo{author}{\bibfnamefont{J.}~\bibnamefont{Britton}},
  \bibinfo{author}{\bibfnamefont{R.~B.} \bibnamefont{Blakestad}},
  \bibinfo{author}{\bibfnamefont{J.}~\bibnamefont{Chiaverini}},
  \bibinfo{author}{\bibfnamefont{D.~B.} \bibnamefont{Hume}},
  \bibinfo{author}{\bibfnamefont{W.~M.} \bibnamefont{Itano}},
  \bibinfo{author}{\bibfnamefont{J.~D.} \bibnamefont{Jost}},
  \bibinfo{author}{\bibfnamefont{C.}~\bibnamefont{Langer}},
  \bibnamefont{et~al.}, \bibinfo{journal}{Nature}
  \textbf{\bibinfo{volume}{438}}, \bibinfo{pages}{639} (\bibinfo{year}{2005}).

\bibitem[{\citenamefont{Anwar et~al.}(2004)\citenamefont{Anwar, Blazina,
  Carteret, Duckett, Halstead, Jones, Kozak, and Taylor}}]{Anwar2004}
\bibinfo{author}{\bibfnamefont{M.~S.} \bibnamefont{Anwar}},
  \bibinfo{author}{\bibfnamefont{D.}~\bibnamefont{Blazina}},
  \bibinfo{author}{\bibfnamefont{H.~A.} \bibnamefont{Carteret}},
  \bibinfo{author}{\bibfnamefont{S.~B.} \bibnamefont{Duckett}},
  \bibinfo{author}{\bibfnamefont{T.~K.} \bibnamefont{Halstead}},
  \bibinfo{author}{\bibfnamefont{J.~A.} \bibnamefont{Jones}},
  \bibinfo{author}{\bibfnamefont{C.~M.} \bibnamefont{Kozak}}, \bibnamefont{and}
  \bibinfo{author}{\bibfnamefont{R.~J.~K.} \bibnamefont{Taylor}},
  \bibinfo{journal}{Phys. Rev. Lett.} \textbf{\bibinfo{volume}{93}},
  \bibinfo{pages}{040501} (\bibinfo{year}{2004}).

\bibitem[{\citenamefont{Baugh et~al.}(2005)\citenamefont{Baugh, Moussa, Ryan,
  Nayak, and Laflamme}}]{Baugh2005}
\bibinfo{author}{\bibfnamefont{J.}~\bibnamefont{Baugh}},
  \bibinfo{author}{\bibfnamefont{O.}~\bibnamefont{Moussa}},
  \bibinfo{author}{\bibfnamefont{C.~A.} \bibnamefont{Ryan}},
  \bibinfo{author}{\bibfnamefont{A.}~\bibnamefont{Nayak}}, \bibnamefont{and}
  \bibinfo{author}{\bibfnamefont{R.}~\bibnamefont{Laflamme}},
  \bibinfo{journal}{Nature} \textbf{\bibinfo{volume}{438}},
  \bibinfo{pages}{470} (\bibinfo{year}{2005}).

\bibitem[{\citenamefont{S{\o}rensen}(1989)}]{Sorensen1989}
\bibinfo{author}{\bibfnamefont{O.~W.} \bibnamefont{S{\o}rensen}},
  \bibinfo{journal}{Prog. Nucl. Magn. Reson. Spectrosc}
  \textbf{\bibinfo{volume}{21}}, \bibinfo{pages}{503} (\bibinfo{year}{1989}).

\bibitem[{\citenamefont{Morris and Freeman}(1979)}]{Morris1979}
\bibinfo{author}{\bibfnamefont{G.~A.} \bibnamefont{Morris}} \bibnamefont{and}
  \bibinfo{author}{\bibfnamefont{R.}~\bibnamefont{Freeman}},
  \bibinfo{journal}{J. Am. Chem. Soc.} \textbf{\bibinfo{volume}{101}},
  \bibinfo{pages}{760} (\bibinfo{year}{1979}).

\bibitem[{\citenamefont{{Durkin} and {Dowling}}(2007)}]{durkin}
\bibinfo{author}{\bibfnamefont{G.~A.} \bibnamefont{{Durkin}}} \bibnamefont{and}
  \bibinfo{author}{\bibfnamefont{J.~P.} \bibnamefont{{Dowling}}},
  \bibinfo{journal}{Phys. Rev. Lett.} \textbf{\bibinfo{volume}{99}},
  \bibinfo{pages}{070801} (\bibinfo{year}{2007}).

\bibitem[{\citenamefont{Wider et~al.}(1984)\citenamefont{Wider, Macura, Kumar,
  Ernst, and Wuthrich}}]{wider}
\bibinfo{author}{\bibfnamefont{G.}~\bibnamefont{Wider}},
  \bibinfo{author}{\bibfnamefont{S.}~\bibnamefont{Macura}},
  \bibinfo{author}{\bibfnamefont{A.}~\bibnamefont{Kumar}},
  \bibinfo{author}{\bibfnamefont{R.~R.} \bibnamefont{Ernst}}, \bibnamefont{and}
  \bibinfo{author}{\bibnamefont{Wuthrich}}, \bibinfo{journal}{J. Mag. Res.}
  \textbf{\bibinfo{volume}{56}}, \bibinfo{pages}{207} (\bibinfo{year}{1984}).

\bibitem[{\citenamefont{Overhauser}(1953)}]{overhauser}
\bibinfo{author}{\bibfnamefont{A.~W.} \bibnamefont{Overhauser}},
  \bibinfo{journal}{Phys. Rev.} \textbf{\bibinfo{volume}{92}},
  \bibinfo{pages}{411} (\bibinfo{year}{1953}).

\bibitem[{\citenamefont{{Cummins} et~al.}(2003)\citenamefont{{Cummins},
  {Llewellyn}, and {Jones}}}]{Cummins2003}
\bibinfo{author}{\bibfnamefont{H.~K.} \bibnamefont{{Cummins}}},
  \bibinfo{author}{\bibfnamefont{G.}~\bibnamefont{{Llewellyn}}},
  \bibnamefont{and} \bibinfo{author}{\bibfnamefont{J.~A.}
  \bibnamefont{{Jones}}}, \bibinfo{journal}{Phys. Rev. A.}
  \textbf{\bibinfo{volume}{67}}, \bibinfo{pages}{042308}
  (\bibinfo{year}{2003}).

\bibitem[{\citenamefont{van Beek}(2007)}]{matNMR}
\bibinfo{author}{\bibfnamefont{J.~D.} \bibnamefont{van Beek}},
  \bibinfo{journal}{J. Magn. Res.} \textbf{\bibinfo{volume}{187}},
  \bibinfo{pages}{19 } (\bibinfo{year}{2007}).

\end{thebibliography}
\end{document}